\documentclass[leqno]{article}
\usepackage{amssymb}
\newcommand{\beq}{\begin{equation}}
\newcommand{\eeq}{\end{equation}}
\newcommand{\ba}{\begin{array}}
\newcommand{\ea}{\end{array}}

\newcommand{\sect}[1]{\setcounter{equation}{0}\section{#1}}

\newcommand{\bea}{\begin{eqnarray}}
\newcommand{\eea}{\end{eqnarray}}
 
\newtheorem{theorem}{Theorem}

\def\b#1{{\mathbb #1}}

\def\nn{\nonumber  \\}

\date{}

\begin{document}
\title{\bf Stability and attractivity for a class of
           dissipative phenomena}
 \author{ {\sc  A. D'Anna   \hspace{30mm} G. Fiore}  \\\\
  Dip. di Matematica e Applicazioni, Fac.  di Ingegneria\\
        Universit\`a di Napoli, V. Claudio 21, 80125 Napoli\thanks{
  This research was supported by Italian Ministry of
  University and Scientific Research.}
       }
 \maketitle

%
%

 {\sc Abstract}: {\small\it We consider initial-boundary-value problems
 for a class of nonlinear third order equations having
 non-autonomous forcing terms and get new asymptotic stability results
 by means of the Liapunov second method.
 The class includes equations arising in Superconductor Theory,
 Quantum Mechanics and in the Theory of Viscoelastic Materials.}

 \vspace{1mm}


 \vspace{1mm}

 {\sc A.M.S. Classification}: {\small 35B35 - 35G30}

\bigskip
\noindent
Preprint 00-39 Dip. Matematica e Applicazioni, Universit\`a di Napoli

\sect{Introduction}

In this paper we study a large class of initial-boundary-value problems
of the form
\beq                               \label{31}
 \:\:\left\{
   \begin{array}{l}
    -\varepsilon u_{xxt}+u_{tt}-c^2u_{xx}= f(x,t,u,u_x,u_{xx},u_t),
    \qquad x\in ]0,1[,\: t>t_0,     \vspace{3mm}\\
    u(0,t)=0, ~~u(1,t)=0
   \end{array}
 \right.
 \eeq
($t_0\ge 0$, $\varepsilon,c$ positive constants), with initial conditions
 \beq                               \label{32}
    u(x,t_0)=u_0(x), ~~u_t(x,t_0)=u_1(x).
\eeq

Many papers \cite{dd1,DacDan98,dr1,dr2,rio,ghi,gr1,gr2}
have been devoted to the analysis of the operator
$L=-\varepsilon \partial_{xxt}+\partial_{tt}-c^2\partial_{xx}$,
which plays a significant role because it characterizes noteworthy
dissipative phenomena. When $f=-b \sin u-au_t+ F(x,t)$, where
$a,b$ are positive constants, we deal with the perturbed
Sine-Gordon  equation related to the classical Josephson effect
in the Theory of Superconductors \cite{dav,lsc}. On the other hand
it is well known \cite{mor} that equation (\ref{31})$_1$  describes
the evolution of the displacement $u(x,t)$ of the section of a rod
from its rest position $x$ in a
Voigt material when an external force $f$ is applied; 
in this case $c^2=E/\rho$, $\varepsilon=1/(\rho\mu)$,
where $\rho$ is the (constant) linear density of the rod at rest,
and $E,\mu$ are respectively
the elastic and viscous constants of the rod, which enter
the stress-strain relation
$\sigma=E\nu+\partial_t \nu/\mu$,
where $\sigma$ is the stress, $\nu$ is the strain.

Now we suppose that (\ref{31}) admits the null solution
$u(x,t)\equiv 0$
and look for new conditions for the stability and attractivity
of the latter, so as to improve some results found in \cite{DacDan98}.
As done there, the distance between the null and a nonnull solution
$u(x,t)$ of the problem  (\ref{31})-(\ref{32})
is introduced as the functional $d(u,u_t)$, where for any
$(\varphi,\psi)\in C^2_0([0, 1])\times C_0([0, 1])$
we define
 \beq                                                \label{33}
  d^2(\varphi,\psi) =
  \int_0^1(\varphi^2+ \varphi_x^2+ \varphi^2_{xx} + \psi^2)dx.
 \eeq
The notions of stability, attractivity  and exponential-asymptotic
stability are formulated using this distance. Imposing the condition
that $\varphi,\psi$ vanish in $0,1$ one easily derives that
$|\varphi(x)|,|\varphi_x(x)|\le d(\varphi,\psi) $ for any $x$;
therefore a convergence in the norm $d$ implies also a
uniform pointwise convergence of  $\varphi,\varphi_x$.

Using  the Liapunov
second method, we first obtain some preliminary
results involving a class of auxiliary functionals depending on a
parameter we will choose according to the examined problem. Supposing
the forcing term satisfies suitable conditions,
we get a theorem of exponential stability. The assumptions we
shall make at the beginning of Section 3 will make the cancellation 
of the term $-c^2 u_{xx}$ by $f$ impossible (this guarantees that 
the nature of the problem cannot be changed by the choice of $f$).
We emphasize that these new hypotheses are now less restrictive 
because they allow the forcing term $f$ to be in a
certain sense unbounded as a function of $t$.
Moreover, this result holds with respect to a metric stronger than $d$,
which can be introduced when the solutions are more regular.
After that, we consider the case that $f$ is specialized as a sum 
of a non-analytic term depending only on $u$ and of
$u_t$ times a bounded function. Suitably modifying the above mentioned 
auxiliary functionals, a theorem of
asymptotic stability in the large is obtained. For each theorem
we give an example of an application.

\sect{Preliminaries}

We shall say that the null solution is

\begin{itemize}

\item {\it uniformly stable} if for any $\varepsilon>0$
there exists a $\delta(\varepsilon)\in ]0,\varepsilon[$ such that
for any $t_0\in J=[0,+\infty[$ the inequality $d(u_0,u_1)<\delta$ implies
$d(u,u_t)<\varepsilon$ for any $t\ge t_0$;

\item {\it attractive} if for any $t_0$
there exist a $\sigma(t_0)>0$ such that
the inequality $d(u_0,u_1)< \sigma$ implies
$d(u,u_t)\rightarrow 0$ as $t\to+\infty$; in particular
{\it attractive in the large} if $\sigma(t_0)=+\infty$ for any $t_0$;

\item {\it uniformly
asymptotically stable (in the large)} if it is uniformly stable and
attractive (in the large);

\item {\it exponential-asymptotically
stable} if there exist constants
$C,D(t_0),\delta'(t_0)>0$ such that
$d(u_0,u_1)<\delta'$ implies
$d(u,u_t)<D \, d(u_0,u_1)e^{-C(t-t_0)}$ for any $t\ge t_0$.

\end{itemize}

As shown in \cite{DacDan98}, a set of sufficient
conditions for the existence and uniqueness of the solution of the problem
(\ref{31}-\ref{32}) in the time interval $[0,T]$ is the following:
\beq                                                           \label{28}
 \mbox{$f(x,t,u,p,q,r)$ is defined and continuous on the set}
 \eeq
 \[
 \{(x,t,u,p,q,r)~|~ 0\leq x \leq 1,~0\leq t \leq T,~-\infty<u,p,q,r<+\infty,
 \  \ T>0 \};
 \]
 \beq                                                           \label{29}
  \mbox{ there exists a constant $\mu>0$ such that}
 \eeq
 \[
    |f(x,t,u_1, p_1,q_1,r_1)-f(x,t,u_2,p_2,q_2,r_2)|
   \leq
 \]
 \[
   \leq \mu\{|u_1-u_2|+|p_1-p_2|+|q_1-q_2|+|r_1-r_2| \};
 \]
 \beq
u_0,~u'_0,~u_0^{''},~u_1~ \mbox{continuous on}~  0\leq x\leq 1\eeq
\[
 \mbox{and such that}~~ u_0(0)=u_0(1)=u_1(0)=u_1(1)=0.
\]
We shall assume that they are all fulfilled for the class
of problems considered in Section \ref{gen3.1}. The
function $f$ considered in Section \ref{gen3.3} does not
satisfy condition (\ref{29}), but however we are able to
obtain the stability properties of the null solution.

To prove our theorems we shall use the Liapunov direct method.
We introduce the Liapunov functional
 \beq                                                 \label{34}
     V(\varphi,\psi)=\frac{1}{2}\int_0^1\{(\varepsilon \varphi_{xx}-\psi)^2+
     \gamma\psi^2+c^2(1+\gamma)\varphi_x^2 \}dx,
 \eeq
 where $\gamma$ is an arbitrary positive constant. It turns out
 that
 \[
  V\leq \frac{1}{2}\int_0^1\{\varepsilon^2 \varphi_{xx}^2 +\psi^2
   +\varepsilon \varphi_{xx}^2 +\varepsilon\psi^2
    + \gamma\psi^2+c^2(1+\gamma)\varphi_x^2 \}dx.
 \]
 Setting

 \beq                                                 \label{35}
     c_2^2= \max \{c^2(1+\gamma)/2, \varepsilon(1+\varepsilon)/2,
           (1+\varepsilon+\gamma)/2\},
 \eeq
 we thus derive

 \beq                                                 \label{36}
     V(\varphi,\psi)\leq c^2_2d^2(\varphi,\psi).
 \eeq
Moreover, it is known that
  \beq                                       \label{37}
  \varphi(0)=0 ~~ \Longrightarrow
  ~~\int_0^1 \varphi^2_x(x)dx\geq \int_0^1 \varphi^2(x)dx,
 \eeq
 and \cite{danr}
 \beq                                       \label{38}
  \varphi(0)=0, ~~\varphi(1)=0 ~~ \Longrightarrow
  ~~\int_0^1 \varphi^2_{xx}(x)dx\geq \int_0^1 \varphi_x^2(x)dx.
 \eeq

 Using (\ref{37}),(\ref{38}) and an argument employed in \cite{DacDan98}, we
 get
  \beq                                       \label{39}
     V(\varphi,\psi)\geq c^2_1d^2(\varphi,\psi).
 \eeq
 where
 \vspace{2mm}
 \beq                                       \label{310}
 c_1^2= \min \{\varepsilon^2/16, c^2(1+\gamma)/2, (\gamma-1/2)/2\},
        ~~(\gamma >1/2),
 \eeq
 Therefore, from (\ref{36}) and (\ref{39}) we find
\beq
\frac V{c_2^2} \le d^2 \le \frac V{c_1^2}.                \label{invemag}
\eeq
On the other hand, choosing $\gamma=1$ in (\ref{34}) it turns out
\bea                                         \label{inter}
 \dot V &=& \int_0^1 \left \{
    -\frac{\varepsilon c^2}{2}u_{xx}^2-\varepsilon u_{xt}^2+
    \frac{\varepsilon}{2}u_{t}^2
       -\frac{\varepsilon}{2}   \left(cu_{xx}+f/c\right)^2 -
 \right. \nn
&& \left. \qquad\qquad
  -\frac{\varepsilon}{2}\left(u_t-2f/\varepsilon \right)^2+
  Af^2 \right \}dx \nn
&\le & -\int_0^1 \left\{\frac{\varepsilon c^2}{6} \left(u^2+u_x^2+
u_{xx}^2\right)+\frac{\varepsilon }{2}  u_t^2+  Af^2 \right\} dx \nn
&\le& -c_3^2 d^2(u,u_t)+  \int_0^1Af^2 dx
\eea
where we have set
\beq
A:=(\varepsilon/2 c^2)+2/\varepsilon, \qquad
c_3^2:=\min\left\{\frac{\varepsilon c^2}{6} ,
   \frac{\varepsilon }{2} \right\},              \label{defs}
\eeq
and we have used (\ref{37}), (\ref{38}). In the sequel we shall
set also $p:=c_3^2/c_2^2$.

\sect{Stability and attraction region for (\ref{31})}
\label{gen3.1}

We introduce the following

\noindent
{\it Hypothesis 1}- Assume that
\beq
A \int_0^1\,f^2 dx\leq \hat g(t,d^2)c_1^2 d^2            \label{hypp}
\eeq
where $f$ is the function of $(\ref{31})$ and
$\hat g(t,\eta)$ ($ t>t_0, \ \eta>0$)
is continuous, nonnegative, non-decreasing in $\eta$
and such that the limit
\beq
\lim\limits_{t\to+\infty}\frac{\int^t_0 \hat
g(\tau,\eta/c_1^2)d\tau}t=:q(\eta)                          \label{hyp2'}
\eeq
defines  a continuous, non-decreasing function
$q:\eta\in J\to J$ with $q(0)<p$.

\bigskip
The assumption that $\hat g(t,\eta)$
is non-decreasing in $\eta$ is no real loss of generality;
if originally this is not the case,
we just need to replace $\hat g(t,\eta)$ by
$\max\limits_{0\le \theta\le \eta}\hat g(t,\theta)$ to fulfill
this condition.

\begin{theorem}
Under these assumptions the null solution of the problem (\ref{31})
is exponential-asymptotically stable and the region of attraction 
related to the initial time
$t_0$ includes the set

 \[
 d(u_0,u_1)< \left[\sup\limits_{r\in ]0,\bar r[} \frac{r}{c_2^2}
 e^{-M(t_0, r)}\right]^{1/2},
 \]
  where $\bar r$ and $M(t_0, r)$
 are defined by (\ref{dino1}) and (\ref{defM}).


\label{theo1}
\end{theorem}

{\sc Proof}.
From (\ref{inter}), using (\ref{hypp}), (\ref{invemag}) and
the monotonicity in $\eta$ of
$g(t,\eta):=\hat g(t,\frac{\eta}{c_1^2})$, we find
\bea
\dot V(u,u_t)
&\le& -c_3 d^2+ \hat g(t,d^2)c_1^2d^2\nn
&\le& [-p+ g(t,V)]V. \nonumber
\eea
By the ``comparison principle''
(Lemma 24.3 of \cite{yos}) $V$ is bound from above
\beq
0\le V(t)\le y(t),             \qquad\qquad\qquad t\ge t_0    \label{ma}
\eeq
by the solution $y(t)$ of the Cauchy problem
\beq
\dot y=[-p+g(t,y)]y, \qquad\qquad y(t_0)=y_0\equiv V(t_0)>0.
                                                    \label{eqconf3}
\eeq
We therefore study the latter.
Problem (\ref{eqconf3}) is equivalent to the integral equation
  \vspace{2mm}
\beq                                           \label{solun}
y(t) = y_0\,e^{- p(t-t_0)+\int^t_{t_0}
g\left(\tau,y(\tau)\right) d\tau}.
\eeq
Let
  \vspace{2mm}
\beq                                             \label{dino1}
\bar r:=\sup\{\rho\ge 0 \:|\: q(\rho)<p\}.
\eeq
The inequality
$q(0)<p$ implies $\bar r>0$. Chosen a $r\in]0,\bar r[$, for
any $\eta\le r$ and $t_0\in J$ condition (\ref{hyp2'}) implies
\bea
\lim\limits_{t\to+\infty}\frac{\int^t_{t_0} g(\tau,\eta) d\tau }{t-t_0}
&\le &\lim\limits_{t\to+\infty}\frac{\int^t_{t_0} g(\tau,r) d\tau }{t-t_0}\nn
&=&\lim\limits_{t\to+\infty}\frac{\int^t_0 g(\tau,r) d\tau -
\int^{t_0}_0 g(\tau,r) d\tau}t\frac t{t-t_0}=q(r), \nonumber
\eea
and therefore $\forall\sigma>0$ $\exists \,t'(\sigma,r,t_0)>t_0$
such that $\forall t>t'$
\[
\frac{\int^t_{t_0} g(\tau,\eta) d\tau }{t-t_0}< q(r)+\sigma.
\]
Choosing $\sigma\equiv\frac{p-q(r)}2$ and, denoting
by $t'_0(r,t_0)$ the corresponding value of $t'(\sigma(r),r,t_0)$, we find that
for any $t>t'_0$, $\eta\le r$
\[
\frac{\int^t_{t_0} g(\tau,\eta) d\tau }{t-t_0}<q(r)+\sigma=\frac{p+q(r)}2,
\]
whence
\beq
~\\[6pt]
- p(t-t_0)+\int^t_{t_0}
g\left(\tau,\eta\right) d\tau < -\frac{p-q(r)}2 (t-t_0). \label{cas1}
\eeq
On the other hand, as $q(r)>0$,
for $t\in]t_0,t'_0]$ and $\eta\le r$
\beq
- p(t-t_0)+\int^t_{t_0}d\tau
g\left(\tau,\eta\right) <
 -\frac{p-q(r)}2 (t-t_0)+M(t_0,r),             \label{cucu}
\eeq
where we have set
 \beq                                        \label{defM}
 M(t_0,r):=\max\left \{0,\max\limits_{t_0\leq t\leq t'_0}
 \left[-\frac{p-q(r)}2 (t-t_0)+\int^t_{t_0} g(\tau,r) d\tau \right]
 \right\}
  \eeq
Looking back at (\ref{cas1}) we realize that
(\ref{cucu}) actually holds for any $t> t_0$,
because by definition $M(t_0,r)\ge 0$.
Therefore, given any $t> t_0$, if the solution of (\ref{eqconf3}) satisfies
$y(\tau)\leq r$ for any $\tau\in[t_0,t[$ then (\ref{solun})
and (\ref{cucu}) imply
\beq
y(t) < y_0\,e^{M(t_0,r)}\,e^{-\frac{p-q(r)}2 (t-t_0)}.
                                                          \label{magg}
\eeq
Now it is easy to show first that, indeed,
\beq
\qquad 0<y_0<r e^{-M(t_0,r)} \qquad 
\Rightarrow\qquad  y(t)<r\qquad \forall t\ge t_0.
                                                   \label{rough}
\eeq
In fact, if {\it per absurdum} there existed $t_1>t_0$ such that
\bea
&&y(\tau;t_0,y_0)<r\qquad\qquad\mbox{for } t_0\le \tau<t_1
\label{cont1}\\
&&y(t_1;t_0,y_0)=r                               \label{cont2}
\eea
then (\ref{magg}), (\ref{cont1}) would imply
\[
y(t_1) < y_0\,e^{M(t_0,r)}\,e^{-\frac{p-q(r)}2 (t_1-t_0)}<
r\,e^{-\frac{p-q(r)}2 (t_1-t_0)}<r,
\]
against (\ref{cont2}). Having proved the bound (\ref{rough}),
now we can immediately improve it.
We can reconsider the first part
of the previous inequality chain based on (\ref{magg})
for {\it any}  $t>t_0$
\[
y(t) < y_0\,e^{M(t_0,r)}\,e^{-\frac{p-q(r)}2 (t-t_0)}
\]
and thus find the implication
\beq
\qquad\quad 0<y_0<r e^{-M(t_0,r)} \qquad \: \Rightarrow\qquad \:
y(t)<y_0\,e^{M(t_0,r)}\,e^{-\frac{p-q(r)}2 (t-t_0)}    \label{fine}
\eeq
for any $t>t_0$. Now (\ref{ma}), (\ref{fine}) imply
\[
V(t) < V(t_0)\,e^{M(t_0,r)}\,e^{-\frac{p-q(r)}2 (t-t_0)} ,
\]
provided $V(t_0)<r e^{-M(t_0,r)}$. With the short-hand notation
$d^2(t)\equiv d^2(u,u_t)$,  we thus find
that the assumption
\beq                                     \label{315}
d^2(t_0)< r \frac{e^{-M(t_0,r)}}{c_2^2}
\eeq
implies, because of (\ref{39}) and (\ref{36}),
 \beq
d^2(t) < d^2(t_0)\,\frac{c_2^2}{c_1^2}e^{M(t_0,r)}
\: e^{-\frac{p-q(r)}2 (t-t_0)},
\eeq
i.e. the exponential-asymptotical stability.
Finally, from (\ref{315}) we derive the attraction region includes
the set
 \beq
 d^2(u_0,u_1)< \sup\limits_{r\in ]0,\bar r[} \frac{r}{c_2^2}e^{-M(t_0, r)}.
 \eeq
 \bigskip

{\bf Remark 1.} This is an alternative to Theorem 3.2 B) of \cite{DacDan98},
which gives sufficient conditions for the
exponential-asymptotical stability of the null solution.
The hypothesis (\ref{hypp}) considered here is much weaker than
the one considered there, where it was required that there exists 
a positive constant $M$ such that
$$
  f^2(x,t,\varphi,\varphi_x,\varphi_{xx},\psi)\leq
  M(\varphi^2+\varphi_x^2+\varphi^2_{xx}+\psi^2),
$$
in that $f$ may well be an unbounded function of $t$
and nonetheless fulfill (\ref{hypp}). This is the case 
for the following 

\bigskip
{\bf Example 1.} Let
$f=b(t)\sin\varphi$, with a function $b(t)$ such that the limit
$\lim_{t\to+\infty}(\int_0^t b^2(\tau)d\tau)/t$ be finite and
smaller than $p$; then we can set $\hat g(t,\eta)\equiv b^2(t)$.
For instance we could take $b^2$ a continuous function that 
vanishes everywhere
except in intervals centered at equally spaced points, where 
it takes linearly increasing maxima but keeps the integral
constant, e.g.
\beq
b^2(t)=b_0\:\left\{
\begin{array}{ll}
n^2(t-n+1/n)\qquad & \qquad\mbox{if} \qquad t\in[n-1/n,n],\cr
n- n^2(t-n) \qquad &\qquad\mbox{if} \qquad t\in ]n,n+1/n],\cr
            0\qquad &\qquad\mbox{otherwise,}
\end{array}\right.
\eeq
with $b_0<p$ and $n=2,3,...$.

\bigskip
{\bf Remark 2.} Under the assumption that the
problem (\ref{31}), (\ref{32}) admits solutions
$u(x,t)$ having also continuous derivative $u_{xtt}$, then
one can replace (\ref{33}) by the metric
\beq                                                \label{nmetric}
d_1^2(\varphi,\psi) =d^2(\varphi,\psi)+\int_0^1 \psi_x^2 dx,
\eeq
(\ref{34}) by the functional
\beq
V_1(\varphi,\psi)=V(\varphi,\psi)+\frac{\varepsilon}{2}\int_0^1
\{\varepsilon\psi_x^2-2c^2\psi\varphi_{xx}\}dx
\eeq
and verify that Theorem \ref{theo1} holds with respect to the metric $d_1$.
 \vspace{4mm}

\sect{Stability and attractivity for a non-analytic forcing term}
\label{gen3.3}

We now specialize the function $f$ of (\ref{31}) as
$f=F(u)-a(x,t,u,u_x,u_t,u_{xx})u_t$, where $F\in C(\b{R})$
and $a\in C(]0,1[\times [0,+\infty[ \times \b{R}^4)$,
and examine the particular problem
  \beq                               \label{320}
 \left\{
   \begin{array}{l}
    Lu=F(u)-a(x,t,u,u_x,u_t,u_{xx})u_t, \qquad x\in ]0,1[,\:\: t>t_0
                                                 \vspace{3mm}\\
    u(0,t)=0, ~~u(1,t)=0,
   \end{array}
 \right.
 \eeq
with initial conditions  (\ref{32}).
We shall use use the modified Liapunov functional
\bea                                               \label{321}
W(\varphi,\psi) &=&
\frac{1}{2}\int_0^1 \left\{(\varepsilon \varphi_{xx}-\psi)^2+
\gamma\psi^2+c^2(1\!+\!\gamma)\varphi_x^2\}dx\right\} \\
&& -(1\!+\!\gamma)\int_0^1
\left(\int_0^{\varphi(x)}F(z)dz\right)dx \nonumber
\eea
where $\gamma>1/2$ for the moment is an unspecified
parameter.

\begin{theorem}
The null solution of the problem
(\ref{320}) is uniformly asymptotically stable in the large
under the following assumptions:
\bea
&&  \mbox{there exist $\tau\in[0,1[$ and $D>0$ such that,
for any $\varphi,\psi$} \label{HYP1} \\
&& 0 \le -\int_0^1\left(\int_0^{\varphi(x)}F(z)dz \right) dx
\le \frac{D}{\gamma+1} d^{\tau+1}(\varphi,\psi); \nn
&&
\int_0^1 F(\varphi(x)) \varphi_{xx}(x)dx \ge 0\ \
 \mbox{for any }\varphi\in C^2_o([0,1]);
 \label{HYP3}\\
&& \mbox{the function $a$ satisfies }\:\:
  \inf a > -\varepsilon, \ \  \sup a < +\infty.      \label{324}
\eea
\end{theorem}

{\sc Proof.} Reasoning as in section 3.4 of reference \cite{DacDan98}, we get
\bea
W(\varphi,\psi) &\ge &
\frac{1}{2}\int_0^1 \{(\varepsilon \varphi_{xx}-\psi)^2+
\gamma\psi^2+c^2(1+\gamma)\varphi_x^2 \}dx \nn
&=& \frac{1}{2}\int_0^1  \{(\varepsilon \varphi_{xx}-2\psi)^2/4+
(\varepsilon \varphi_{xx}-\psi)^2/2+(\gamma-1/2)\psi^2 \nn
&& \qquad+ c^2(1+\gamma)\varphi^2_x+\varepsilon^2 \varphi_{xx}^2/4  \}dx \nn
&\ge&\frac{1}{2}\int_0^1  \{(\gamma-1/2)\psi^2+c^2(1+\gamma)
\varphi^2_x+\varepsilon^2 \varphi_{xx}^2/4  \}dx \nn
&\ge&\frac{1}{2}\int_0^1  \{(\gamma-1/2)\psi^2+c^2(1+\gamma)
(\varphi^2+\varphi^2_x)/2+\varepsilon^2 \varphi_{xx}^2/4  \}dx \nn
&\ge& k_1 d^2(\varphi,\psi),                             \label{lalla}
\eea
where
\beq
k_1:=\frac 12\min\left\{\gamma-\frac 12,\frac{\varepsilon^2}4,
\frac{c^2(1+\gamma)}2 \right\}.
\eeq

Moreover, taking the derivative of $W$
and reasoning as we have done for (\ref{inter}), we obtain
 \[
     \dot W(u,u_t)=-\int_0^1 \{c^2\varepsilon u^2_{xx}+
   \varepsilon\gamma u^2_{xt}+a(1+\gamma)u_t^2+\varepsilon
   F(u)u_{xx}-\varepsilon a u_{xx} u_t  \}dx.
 \]
 From this, considering inequalities (\ref{37}), (\ref{38}) it follows
 \bea                                             \label{322}
  \dot W(u,u_t) &\leq & -\int_0^1 \{(3/4)c^2\varepsilon u^2_{xx}+
   [\varepsilon\gamma+a(1+\gamma-\varepsilon a/c^2 )]u_t^2+ \\
   &&\qquad\varepsilon F(u)u_{xx} + \varepsilon [(c/2)u_{xx}-(a/c)u_t]^2
     \}dx. \nonumber
 \eea
Because of (\ref{HYP3}), the third and fourth terms at the right-hand side are
nonnegative and therefore by (\ref{37}), (\ref{38})
\bea
\dot W(u,u_t) &\leq& -\int_0^1 \{(3/4)c^2\varepsilon u^2_{xx}+
[\varepsilon\gamma+a(1+\gamma-\varepsilon a/c^2 )]u_t^2\}dx \nn
&\le & -\int_0^1 \{c^2/4\varepsilon (u^2_{xx}+u^2_x+u^2)+
[\varepsilon\gamma+a(1+\gamma-\varepsilon a/c^2 )]u_t^2\}dx. \nonumber
\eea
Owing to (\ref{324}) we choose
\[
\gamma=[1+\sup \left\vert a(a\varepsilon/c^2-1)\right\vert]/
(\varepsilon+\inf a) + \frac 12,
\]
so that the coefficient of $u_t^2$ at the right-hand side becomes $\ge 1$ and
we find
\beq
\dot W(u,u_t)\le -k_3 d^2(u,u_t),                     \label{wp}
\eeq
where $k_3:=\min\{c^2/4\varepsilon,1\}$.

Finally, taking into account formula (\ref{321}), assumption (\ref{HYP1}),
 and noting that 
$(\varepsilon \varphi_{xx}-\psi)^2\le\varepsilon^2 \varphi_{xx}^2+
\psi^2+\varepsilon(\varphi_{xx}^2+\psi^2)$,
it follows
\beq                                                 \label{328}
W(\varphi,\psi) \leq c_2^2d^2(\varphi,\psi)+
D d^{\tau+1}(\varphi,\psi).
\eeq
Hence we find
\beq
d^2\ge \min\left\{\frac W{2c_2^2}, \left(\frac W{2D}\right)^{\frac 2{\tau+1}}
\right\},
\eeq
which considered in (\ref{wp}) gives
\beq
\dot W(u,u_t)\le -k_3 \min\left\{\frac W{2c_2^2},
\left(\frac W{2D}\right)^{\frac 2{\tau+1}}\right\}\le 0.
                                                     \label{wpp}
\eeq
The right-hand side is smaller than zero for any
nonnull choice of the initial conditions, what we shall assume
in the sequel.
By (\ref{wpp}) $W$ is a decreasing function of $t$. This implies
in particular that $W(t)< W(t_0)$ for $t> t_0$, whence,
by (\ref{lalla}), (\ref{328}),
\beq
d^2(t)\le \frac{W(t)}{k_1}<\frac{W(t_0)}{k_1}\le 2
\max\{c_2^2d^2(t_0), D d^{\tau+1}(t_0)\},
\eeq
implying the uniform stability of the null solution.

We now show that for any choice of the
initial conditions $W$ decreases to zero (at least) as a power
of $t$ for $t\to +\infty$. If at $t=t_0$
\beq
\frac W{2c_2^2} \ge \left(\frac W{2D}\right)^{\frac 2{\tau+1}},
\eeq
then, by the monotonicity of $W(t)$, for all $t\ge t_0$
this will be true and (\ref{wpp}) will become
\beq
\dot W(u,u_t)\le -k_3 \left(\frac W{2D}\right)^{\frac 2{\tau+1}};
                                             \label{wppp}
\eeq
by the comparison principle
it will follow
\[
W(t)\le y(t),
\]
where $y(t)$ is the solution of the Cauchy problem
\[
\dot y(t)= -k_3 \left(\frac y{2D}\right)^{\frac 2{\tau+1}}\:, \qquad\quad
y(t_0)=W(t_0),
\]
namely
\beq
W(t)\le y(t)=\frac 1{[W(t_0)+E(t-t_0)]^{\frac{1+\tau}{1-\tau}}},
\eeq
where $E:=\frac {k_3}{(2D)^{\frac 2{\tau+1}}}\frac{1-\tau}{1+\tau}>0$.
Clearly $y(t)$ is decreasing and goes to zero as
$1/t^{\frac{1+\tau}{1-\tau}}$ when $t\to+\infty$.
Recalling (\ref{lalla}), we find
\beq
d^2(t) \le \frac{W(t)}{k_1}\le
\frac 1{k_1[W(t_0)+E(t-t_0)]^{\frac{1+\tau}{1-\tau}}}
                                            \label{final1}
\eeq
for $t\ge t_0$, implying the attractivity
of the null solution in this case.
If on the contrary
\[
\frac {W(t_0)}{2c_2^2} < \left(\frac {W(t_0)}{2D}\right)^{\frac 2{\tau+1}},
\]
(\ref{wpp}) will imply for some time
\[
\dot W(u,u_t)\le -k_3  W
\]
and by the comparison principle an (at least) exponential decrease
of $W$.  Hence there will
exist a $T>t_0$ such that
\[
\frac{W(T)}{2c_2^2} = \left(\frac {W(T)}{2D}\right)^{\frac 2{\tau+1}},
\]
after which (\ref{wpp}) will take again the form (\ref{wppp}) and thus imply
\beq
d^2(t) \le \frac{W(t)}{k_1}
\le \frac 1{k_1[W(T)+E(t-T)]^{\frac{1+\tau}{1-\tau}}}
\eeq
for $t\ge T$. This implies the attractivity in the large
 of the null solution.

\bigskip
{\bf Example 2}. An example of a forcing term
fulfilling the conditions (\ref{HYP1}), (\ref{HYP3}) is
the non-analytic one
\beq
F(u)=-k\, \mbox{sign}(u) |u|^{\tau}\qquad 0< \tau\le 1, \:\:
k=\mbox{const}>0.
\eeq
In fact, in this case (\ref{HYP1}) is fulfilled since
\[
-\int_0^1\left(\int_0^{\varphi(x)}F(z)dz\right) dx
=\frac k{\tau+1}\int_0^1 |\varphi(x)|^{\tau+1}dx \ge 0,
\]
and,  by Schwarz inequality and (\ref{37}), (\ref{38})
\bea
&&\int_0^1|u|^{\tau+1}dx\le
\left(\int_0^1u^2dx \right)^{\frac{\tau+1}2} \nn
&&\le \left(\frac 13 \int_0^1(u^2+u^2_x+u^2_{xx})dx
\right)^{\frac{\tau+1}2}\le \frac 1{3^{\frac{\tau+1}2}}d^{\tau+1}(u,u_t);
\nonumber
\eea
(\ref{HYP3}) is fulfilled since, integrating by parts,
\beq
\qquad \int_0^1 F(u) u_{xx}(x)dx
=-k\int_0^1 \mbox{sign}(u) |u|^{\tau}
u_{xx}(x)dx=\tau k\int_0^1 \frac{u_x^2}{|u|^{1-\tau} }
\ge 0. \nonumber
\eeq

 \vspace{4mm}

{\bf Remark 3.}
This result should be compared with Thm 3.3. in reference
\cite{DacDan98}: the claim is the same, but
the hypotheses are adapted to cover
the case of a non-analytic forcing term.

\end{document}